\documentclass[%
 aip,
 amsmath,amssymb,
preprint,%
]{revtex4-1}

\usepackage{graphicx}
\usepackage{dcolumn}
\usepackage{bm}

\usepackage[utf8]{inputenc}
\usepackage[T1]{fontenc}
\usepackage{mathptmx}
\usepackage{etoolbox}
\usepackage{wasysym}
\makeatletter
\def\@email#1#2{%
 \endgroup
 \patchcmd{\titleblock@produce}
  {\frontmatter@RRAPformat}
  {\frontmatter@RRAPformat{\produce@RRAP{*#1\href{mailto:#2}{#2}}}\frontmatter@RRAPformat}
  {}{}
}%
\makeatother
\begin{document}

\preprint{AIP/123-QED}

\title[Loading of the narrow line Tm MOT from a pulsed cold atomic beam]{Loading of the narrow line Tm MOT from a pulsed cold atomic beam}
\author{M. Yaushev}
\email{iaushev.mo@gmail.com}
 \affiliation{P.N.\,Lebedev Physical Institute, Leninsky prospekt 53, 119991 Moscow, Russia}
 \affiliation{Russian Quantum Center, Bolshoy Bulvar 30,\,bld.\,1, Skolkovo IC, 121205 Moscow, Russia}

\author{D. Mishin}

 \affiliation{P.N.\,Lebedev Physical Institute, Leninsky prospekt 53, 119991 Moscow, Russia}
 \affiliation{Russian Quantum Center, Bolshoy Bulvar 30,\,bld.\,1, Skolkovo IC, 121205 Moscow, Russia}

\author{D. Tregubov}

 \affiliation{P.N.\,Lebedev Physical Institute, Leninsky prospekt 53, 119991 Moscow, Russia}
 \affiliation{Russian Quantum Center, Bolshoy Bulvar 30,\,bld.\,1, Skolkovo IC, 121205 Moscow, Russia}

\author{N. Kolachevsky}

 \affiliation{P.N.\,Lebedev Physical Institute, Leninsky prospekt 53, 119991 Moscow, Russia}
 \affiliation{Russian Quantum Center, Bolshoy Bulvar 30,\,bld.\,1, Skolkovo IC, 121205 Moscow, Russia}

\author{A. Golovizin}%
 \email{artem.golovizin@gmail.com}
 \affiliation{P.N.\,Lebedev Physical Institute, Leninsky prospekt 53, 119991 Moscow, Russia}
 \affiliation{Russian Quantum Center, Bolshoy Bulvar 30,\,bld.\,1, Skolkovo IC, 121205 Moscow, Russia}
 
\date{\today}

\begin{abstract}
We report on building a pulsed source of cold Tm atoms and loading of the narrow-line magnito-optical trap (MOT) from the cold atomic beam. 
We achieve the loading rate of the first-stage MOT in the primary chamber up to $10^8$ atoms/s and obtain a cold atomic beam with the mean longitudinal velocity $\sim10$\,m/s and angular spread of 18\,mrad in a pulsed mode.
We also introduce a novel method to enhance the capture velocity of the narrow-line MOT by incorporating additional axial cooling beam, and achieve loading efficiency $\eta = 10\%$ of the second-stage MOT in the science chamber. 
Our approach could be extended to other atomic species with similar properties, like Sr, Yb, Dy and Er, serving as a convenient alternative for the traditional 2D-MOT schemes.
Providing rapid loading of the MOT in the science chamber, it would reduce preparation time of the atomic ensemble leading to a shorter dead time in spectroscopy experiments and higher repetition rate. 
\end{abstract}

\maketitle

\section{Introduction}

Ensembles of ultracold atoms have become a popular platform for many experiments in the fields of quantum sensing \cite{bongs2019taking}, quantum metrology \cite{RevModPhys.87.637}, and quantum simulation \cite{Bloch2012}. With rapid progress in the field of atomic physics, a lot of attention goes to experiments with open-f-shell lanthanides such as Er, Dy, and Tm.
These atoms possess complex energy structures that offer a diverse range of electronic transitions, like those in Sr and Yb, along with unique properties tied to their open 4f shell, such as a large effective spin and magnetic moment.
This makes lanthanides a fascinating platform for many experiments, e.g., they offer opportunities to explore systems with long-range interactions \cite{chomaz2022dipolar}.

An important characteristic of lanthanides is their high melting points at $1400-1550\,^\circ$C, which leads to the necessity of operating an atomic oven at more than $700\,^\circ$C. 
This has historically necessitated the use of Zeeman slower setups \cite{PhysRevLett.48.596}, as the capture velocity in magneto-optical traps (MOTs) rarely exceeds 50\,m/s even for broad transitions, which is significantly lower than the thermal velocity of atomic beams emitted from ovens.
Consequently, these setups tend to be large and low-efficient, with a small fraction of the atomic flux successfully loaded into the remote MOT.
Over the past decade, two-dimensional magneto-optical traps (2D-MOTs) have become a common and efficient alternative to traditional Zeeman slower systems. 
Initially demonstrated with Rb \cite{PhysRevA.58.3891}, this approach has since been successfully adapted for various alkali and alkaline-earth elements, including Li \cite{PhysRevA.80.013409}, Na \cite{Lamporesi_2013}, K \cite{10.1063/5.0154985}, Sr \cite{PhysRevA.96.053415, Kwon_2023}, and Yb \cite{D_rscher_2013, saskin2019narrow}. 
2D-MOT setups are not only compact and efficient due to the close proximity of the source of hot atoms to the 2D-MOT center, but they also facilitate reaching ultrahigh vacuum due to the spatial separation of the hot-atom oven and the science chamber.
The latter is particularly advantageous for experiments involving Bose–Einstein condensates (BEC) and quantum information, where long atom lifetimes in dipole traps are crucial.
Additionally, 2D-MOTs can provide a continuous flux of cold atoms, which could enable the development of a new generation of optical clocks that continuously interrogate a clock transition \cite{katori2021longitudinal, mishin2021continuous, dubey2024modelingcontinuoussuperradiantlaser}. This approach helps to suppress the Dick effect \cite{Dick, 710548}, which often limits stability of the optical clocks.

Tm is a lanthanide with only one vacancy in the 4f shell which gives it a moderate magnetic moment of $\mu = 4\,\mu_b$ in comparison to Er (7\,$\mu_b$)  or Dy (10\,$\mu_b$), where $\mu_b$ is the Bohr magneton. 
On the other hand, Tm's simpler energy structure and less dense Feshbach resonance spectrum, comparing to other open-f-shell lanthanides, make it a promising candidate for quantum simulation \cite{davletov2020machine}. 
Furthermore, Tm is an excellent fit for transportable optical lattice clock due to the low sensitivity to black-body radiation \cite{golovizin2019inner}, convenient magic wavelength for optical lattice formation \cite{sukachev2016inner}, and the synthetic frequency technique allowing to eliminate the second order of the Zeeman shift \cite{golovizin2021simultaneous}.

Despite the widespread adoption of 2D-MOT setups for alkali and alkaline-earth elements, loading of a narrow-line MOT from a cold atomic beam for open-f-shell elements has been demonstrated only for Dy \cite{PhysRevA.108.023719}. 
Here we report on the development of a pulsed source of cold Tm atoms, which is based on a first-stage MOT in a primary chamber, and the loading of the narrow-line MOT from the cold atomic beam in the separate science chamber.
We introduce an innovative approach to enhance the capture velocity of the secondary MOT by leveraging a dual-chamber design.

The manuscript is organized as follows: the Sec.\,\ref{sec:vacuum} gives an overview of Tm energy level structure and describes the vacuum apparatus and optical part of the experiment. 
Sec.\,\ref{sec:blue_mot} is dedicated to characterizing the first-stage cooling performance. 
In the Sec.\,\ref{sec:atomic_beam} we investigate properties of the cold atomic beam generated from the primary chamber. 
Next, Sec.\,\ref{sec:green_mot} presents the results of loading the narrow-line MOT in the science chamber. 
Finally, we summarize our findings in the Sec.\,\ref{sec:end}.

\section{Experimental setup}\label{sec:vacuum}

\begin{center}
\begin{figure}
\includegraphics[width=\linewidth]{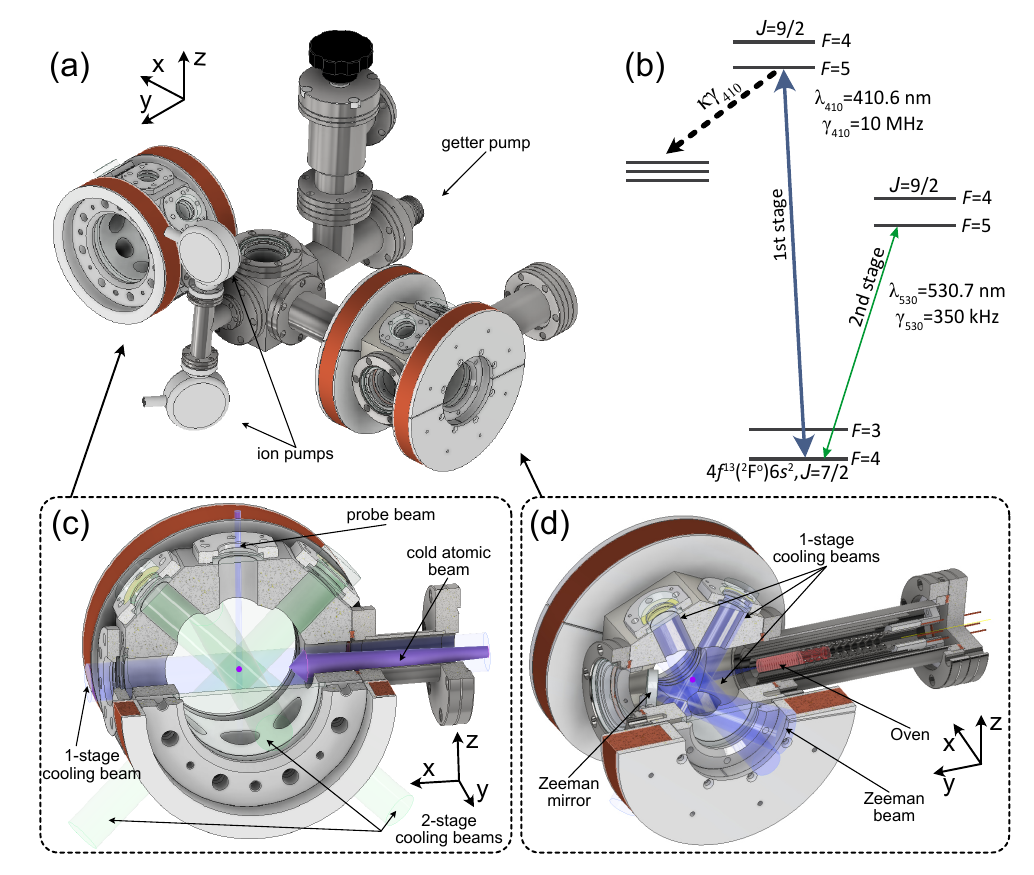}
\caption{\label{fig:vacuum} Experimental setup. a) Vacuum system CAD drawing. 
MOT coils are shown with copper color. 
b) Simplified Tm level structure, dashed line indicates decay channels from upper level (see main text).
c) Detailed view of the science chamber with cooling and probe beams (transparent blue and green cylinders) and cold atomic beam (light purple arrow). 
d) Detailed view of the primary chamber. 
Atomic oven consists of a sapphire cylinder (shown with red), still tube holder and thermal shields.
All laser beams (except Zeeman beam) are retro-reflected. }
\end{figure}
\end{center}

Simplified level diagram of Tm with the relevant transitions is depicted on Fig.\,\ref{fig:vacuum}.b. 
For the first stage cooling 
we use the broad 410.6\,nm transition $
\left|4 f^{13}\left({ }^2 F^o\right) 6 s^2, J=7 / 2, F=4\right\rangle \rightarrow\left|4 f^{12}\left({ }^3 H^5\right) 5 d_{3 / 2} 6 s^2, J=9 / 2, F=5\right\rangle
$ with a natural linewidth of 10\,MHz, which correponds to the Doppler temperature limit of $200$\,\textmu K.
We produce up to 120\,mW of 410.6\,nm radiation using a home-built bow-tie second harmonic generation cavity \cite{Shpakovsky2016} and a semiconductor laser with a tapered amplifier
(Sacher Lasertechnik Group) with up to 700\,mW output power at 821\,nm.
The laser frequency is stabilized to a ``master'' semiconductor laser at 821\,nm, which is phase-coherently locked to an active-H-maser-referenced optical frequency comb. 

Second cooling stage is realized using a  350\,kHz-natural-linewidth transition $\left|4 f^{13}\left({ }^2 F^o\right) 6 s^2, J=7 / 2, F=4\right\rangle \rightarrow$ $\left|4 f^{12}\left({ }^3 H^6\right) 5 d_{5 / 2} 6 s^2, J=9 / 2, F=5\right\rangle$ at the wavelength
530\,nm and has a Doppler temperature limit of $9$\,\textmu K. 
This radiation is  obtained by frequency doubling of 1061\,nm ECDL laser (TOptica), which is stabilized to a high finnese ($\mathcal{F} = 150000$) ULE cavity.

The sketch of the dual chamber vacuum apparatus is shown in Fig.\,\ref{fig:vacuum}. 
The distance between the chambers' centers is $L=34.5$\,cm.
Hereafter, we refer to the vacuum chamber which is connected to the atomic oven as a ``primary'' chamber and the other one as a ``science'' chamber.

The primary chamber (Fig.\,\ref{fig:vacuum}.d) is inspired by our previous compact setup \cite{golovizin2021compact}, but has the following major differences: (1) it has a hexagonal shape, (2) it is made from titanium and (3) all optical windows are made of fused silica with antireflection coating in $400-700$\,nm range, which are indium-sealed directly on the chamber.
The atomic oven is connected with standard CF-40 flange at the distance of 8\,cm from the MOT center. 
Such design allows to build a compact and quite efficient source of cold atoms.
Two pairs of MOT laser beams go through $\diameter=19$\,mm windows at an angle of $64^\circ$ to the direction of atomic beam from the oven.
Additional $\diameter=50$\,mm window is used to (1) deliver ``Zeeman'' cooling beam, which is reflected towards the atomic beam along the Y-axis with an in-vacuum mirror, (2) optional third MOT cooling beam, which goes through both chambers along the X-axis and is used in this work to enhance capture velocity of the second-stage MOT in science chamber (see Sec.\,\ref{sec:green_mot}), and (3) probe beam, which is aligned with the third MOT beam. 

The science chamber (Fig.\,\ref{fig:vacuum}.c) has a shape of an octagon (also made from titanium) with seven $\diameter=19$\,mm and two $\diameter=71$\,mm windows, which are also indium-sealed. 
This geometry provides good optical access for all the necessary laser beams (cooling, probe, etc.), while at the same time allows installation of an objective with NA up to 0.7.

The primary and science chambers are connected through a CF-40 cube, CF25-CF40 adatper and 125-mm-long CF-40 tube, which in future may be equipped with a differential vacuum stage.
We use a 140\,l/s non-evaporating getter pump together with two 3\,l/s ion pumps, providing a vacuum level at $10^{-8} - 10^{-9}$\,mbar.

\section{First-stage MOT in the primary chamber}\label{sec:blue_mot}

The loading rate of the MOT in the primary chamber sets an upper limit for a loading rate of the narrow-line MOT in the science chamber.
In addition to mentioned above, the following modifications have been made to the design of the primary chamber compared to the compact setup \cite{golovizin2021compact}: the cooling beams in the new setup goes at an angle $64^\circ$  instead of $45^\circ$, the atomic oven is placed 2\,cm closer to the MOT center and  the Zeeman beam goes at the different angle owing to a more compact design of the chamber. 
Taking this into account, it is necessary to perform optimization of the experimental parameters to achieve the highest loading rate of the first-stage MOT. 
The detailed sketch of the primary chamber is shown on Fig.\,\ref{fig:vacuum}.d.

We use three acousto-optical modulators (AOM) to form MOT cooling beams, Zeeman beam and near-resonant probe beam, respectively. 
While the laser system can deliver up to 120\,mW of 410.6\,nm radiation, we typically operate it at 80\,mW, where its performance is stable. 
We split the output with two non-polarising beamsplitters as follows: 40\,mW goes for MOT beams, 28\,mW goes to the Zeeman beam, and rest (12\,mW) goes to the probe beam.
The MOT beams has $1/e^2$ radius of 6\,mm. 
The Zeeman beam is expanded to the diameter of $\sim25$\,mm and than is focused on the atomic oven using $f = 150$\,mm plano-concave lens.
This allows to maximize the solid angle in which atoms in atomic beam are decelerated by the Zeeman beam.
The probe beam has a  2\,mm diameter, which ensures imaging of the atomic cloud with a maximum size $\sim1$\,mm at the full saturation.

\subsection{Cooling beams detuning optimization}
\begin{figure}[h!]
\includegraphics[width = 0.5\linewidth]{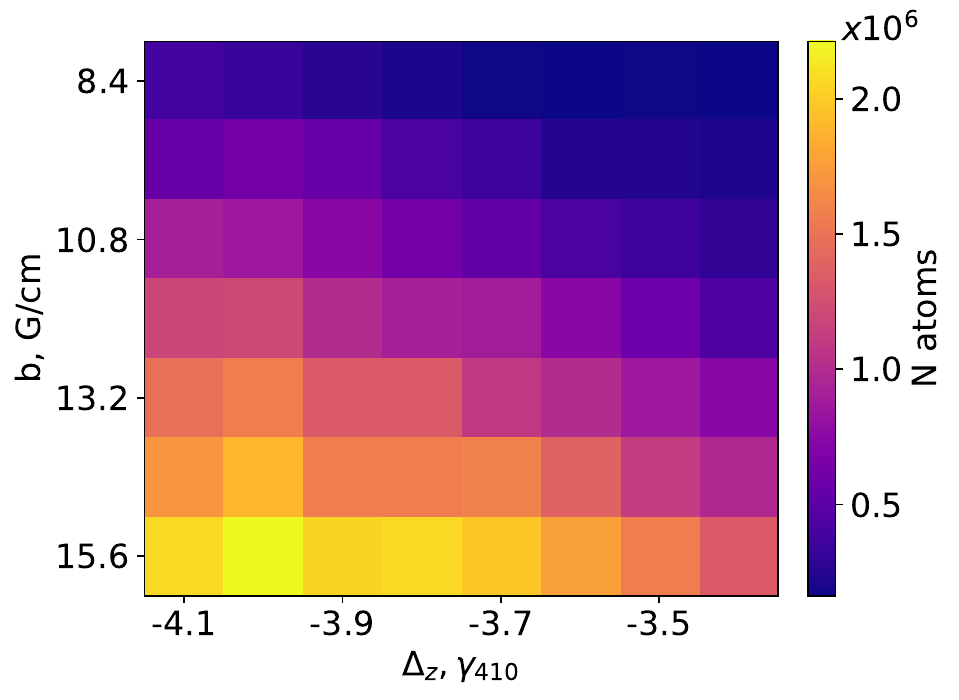}
\caption{\label{fig:zeeman} Experimentally measured number of atoms in millions in the 7-beam MOT in primary chamber as a function of the Zeeman beam detuning (in units of $\gamma_{410}=10$\,MHz) and magnetic field gradient ($6$\,(G/cm)/A).}
\end{figure}

As we demonstrated in the previous compact setup\cite{golovizin2021compact}, such  7-beam MOT configuration works in a wide range of cooling beams detunings and is the most sensitive to the alignment of the Zeeman beam and the hot atomic flux, as well as to the polarization and detuning of the Zeeman beam.
We start with a quick scan of the MOT beams detuning and found the best performance at $\Delta_\textrm{mot} = -1.3\,\gamma_{410} = -13$\,MHz.
Then we proceed with maximizing the number of atoms in the MOT by adjusting the Zeeman beam angle and polarization.
Finally, we performed a scan of the Zeeman beam frequency detuning and magnetic field gradient as it is shown in Fig. \ref{fig:zeeman}. 
The optimum Zeeman beam detuning is $\Delta_\textrm{z} = -4.0\,\gamma_{410} = -40$\,MHz  at magnetic field gradient of $13$\,G/cm along the atomic beam direction.
This value of the magnetic field gradient is the maximum achievable in the current setup configuration.
This initial optimization was done at the atomic oven temperature $T = 440\,^\circ$C.

\subsection{Cooling power dependencies}

The dynamics of the number of atoms in the MOT can be described by the following differential equation:

\begin{equation}\label{eq:effective_lifetime}
    \frac{\mathrm{d}N}{\mathrm{d}t} = R - N(\Gamma^l_\textrm{coll} + \Gamma^l_\textrm{410}),
\end{equation}
where we neglected two-body losses owing to relatively small density of atoms achieved in our MOT. 
Here $R$ is the MOT loading rate, and the second term describes different loss channels. $\Gamma^l_\textrm{coll}$ reflects total losses due to collisions with background buffer gas $\Gamma^l_\textrm{bg}$ and hot atomic beam $\Gamma^l_\textrm{F} \propto F$, where $F$ is the flux of thulium atoms from the oven.
Both these loss rates depend on the oven temperature.
$\Gamma^l_\textrm{410}$ originates from a finite probability $\kappa=3\times10^{-7}$ for an atom to decay from the upper cooling level to some metastable state \cite{Kolachevsky_2007, Phd_Sukachev} (see Fig.\,\ref{fig:vacuum}.b), causing it to exit the cooling cycle. 
Thus, $\Gamma^l_\textrm{410}$ depends on the cooling beam intensity as
\begin{equation} \label{eq:gamma_laser}
    \Gamma^l_\textrm{410} = \frac{\kappa\Gamma_{410}}{2}  \frac{s}{1+s+4(\Delta/\Gamma_{410})^2}
\end{equation} 
where $s=I/I_\textrm{sat}$ is the saturation parameter ($I_\textrm{sat}=18\,\textrm{mW}/\textrm{cm}^2$ is the saturation intensity for 410\,nm transition)  and $\Delta$ is the detuning of radiation from the resonance.
Below we distinguish $\Gamma^\textrm{l,mot}_{410}$ associated with 6 MOT beams and $\Gamma^\textrm{l,z}_{410}$ from Zeeman beam.

\begin{figure}[h!]
\includegraphics[width=0.5\linewidth]{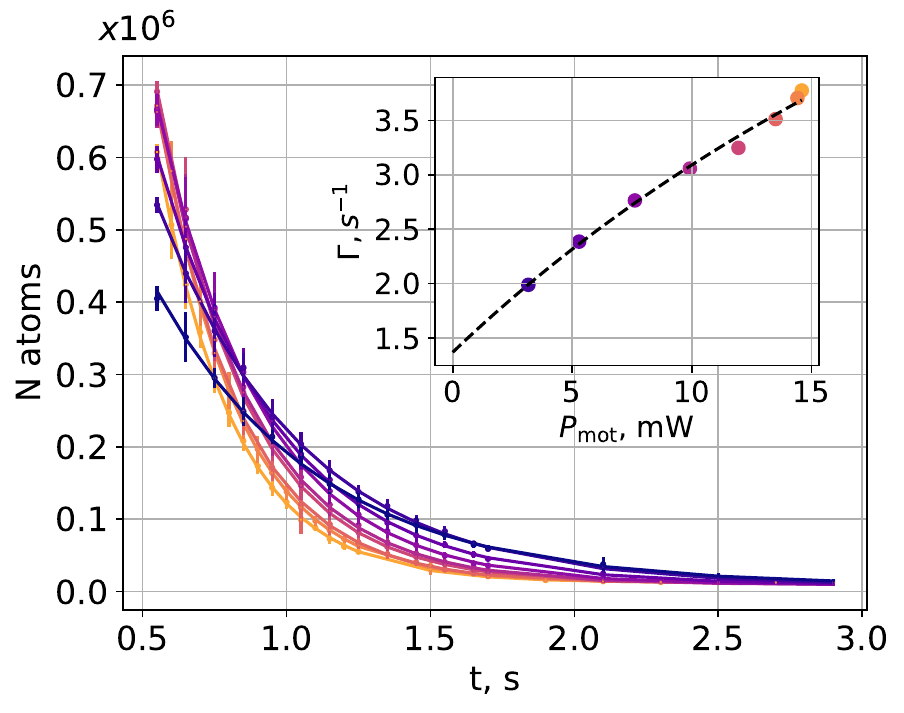}
\caption{\label{fig:decay_blue_mot} 
Number of atoms as a function of 6 beam MOT power. 
Inset shows fit to the Eq.\,\ref{eq:gamma_coll}.} 
\end{figure}

We start with the measurement of atoms lifetime in the first-stage MOT as a function of total power $P_\textrm{mot}$ of 6 MOT beams (see  Fig.\,\ref{fig:decay_blue_mot}). 
For this we perform loading in the configuration described above and then switch off Zeeman cooling beam, which stops loading of atoms (in terms of Eq.\ref{eq:effective_lifetime} $R=0$). 
After 50\,ms we change the power of MOT beams to a desired value and record decay of the number of atoms in the trap. 
Each data set we fit with exponential decay function $N(t) = N_0 e^{-t/\Gamma^l}$.
Inferred values of $\Gamma^l$ are shown in the inset of 
Fig.\,\ref{fig:decay_blue_mot}; the dashed line depicts approximation of this data with 
\begin{equation}
\label{eq:gamma_coll}
\begin{split}
    \Gamma^l(P_\textrm{mot}) &= \Gamma^l_\textrm{coll} + \Gamma^\textrm{l,mot}_{410}(P_\textrm{mot}) \\
    &= \Gamma^l_\textrm{coll} + \frac{\kappa\Gamma_{410}}{2}  \frac{s_\textrm{mot} }{1+s_\textrm{mot} + 4(\Delta_\textrm{mot}/\Gamma_{410})^2}
\end{split}
\end{equation} 
with two fit parameters $\Gamma^l_\textrm{coll}$ and $\eta_\textrm{mot}$ in $s_\textrm{mot} = \eta_\textrm{mot}\frac{2P_\textrm{mot}}{\pi\sigma^2 I_\textrm{sat}}$ to account for possible systematic error in intensity determination, $\sigma=6$\,mm at $1/e^2$ intensity level.
We obtain $\Gamma^l_\textrm{coll}=1.4(0.1)\,\textrm{s}^{-1}$ and $\eta_\textrm{mot}=1.1(0.1)$. 

\begin{figure}[h!]
\includegraphics[width = 0.7\linewidth]{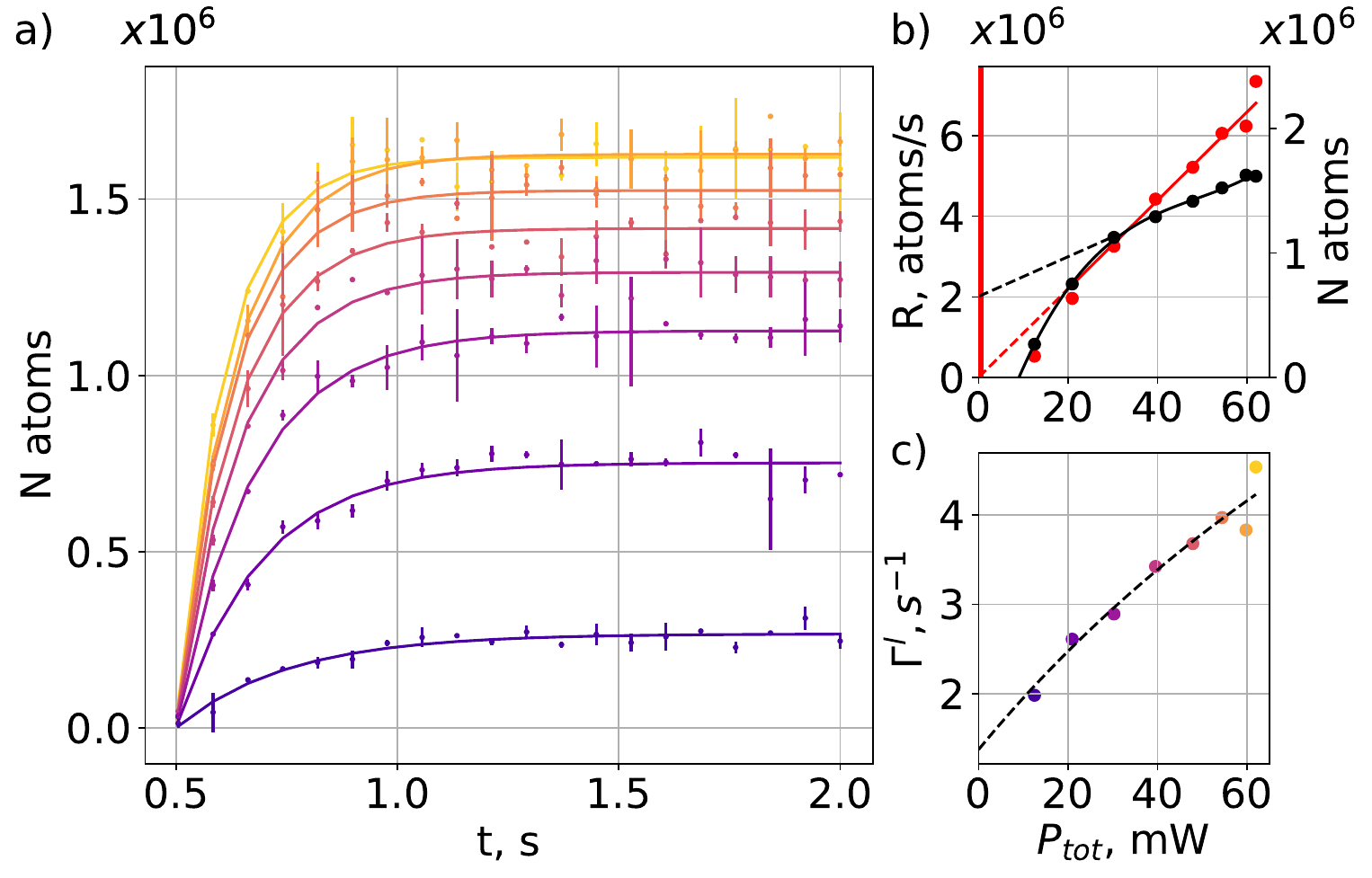}
\caption{\label{fig:load_times} Loading 7-beam MOT at different total laser powers $P_\textrm{tot}$. 
a) MOT atom number as a function of time for different powers (corresponding colors are shown in (c)). 
b)  Loading rate (left axis, red circles) and equilibrium number of atoms (right axis, black circles) as a function of cooling laser power. 
c) Fit of the extracted loss rates to the Eq.\,\ref{eq:gamma_load}.}
\end{figure}

Next, we measure the MOT loading dynamics for different total cooling laser power $P_\textrm{tot}$ (see Fig.\,\ref{fig:load_times}.a), which is divided in the  proportion $10:7$ between MOT and Zeeman cooling beams, respectively.
We fit the loading curves with
\begin{equation}
\label{eq:gamma_load}
\begin{split}
N(t) &= \frac{R}{\Gamma^l} (1-e^{-t\Gamma^l}) \\
\Gamma^l &= \Gamma^l_\textrm{coll} + \Gamma^\textrm{l,mot}_{410}(P_\textrm{tot}) + \Gamma^\textrm{l,z}_{410}(P_\textrm{tot})
\end{split}
\end{equation}
where we used the values of $\eta_\textrm{mot}$ and $\Gamma^l_\textrm{coll}$ from the above, leaving $\eta_\textrm{z}$ (defined similar to $\eta_\textrm{mot}$) as the only free parameter.
Dependence of $\Gamma^l$ on the total cooling laser power is shown on Fig.\,\ref{fig:load_times}.c with dashed line indicating fit to the data with $\eta_\textrm{z}=1.1(0.3)$.
In Fig.\,\ref{fig:load_times}.b one can see an almost linear dependence of the loading rate $R$ for $P_\textrm{tot}>20$\,mW and a continuous growth of the equilibrium number of atoms $N$ in the MOT, which implies the possibility to further increase the number of atoms with higher cooling laser power. 

\subsection{Oven temperature dependencies}
\begin{figure}[h]
\begin{center}
\includegraphics[width=0.7\linewidth]{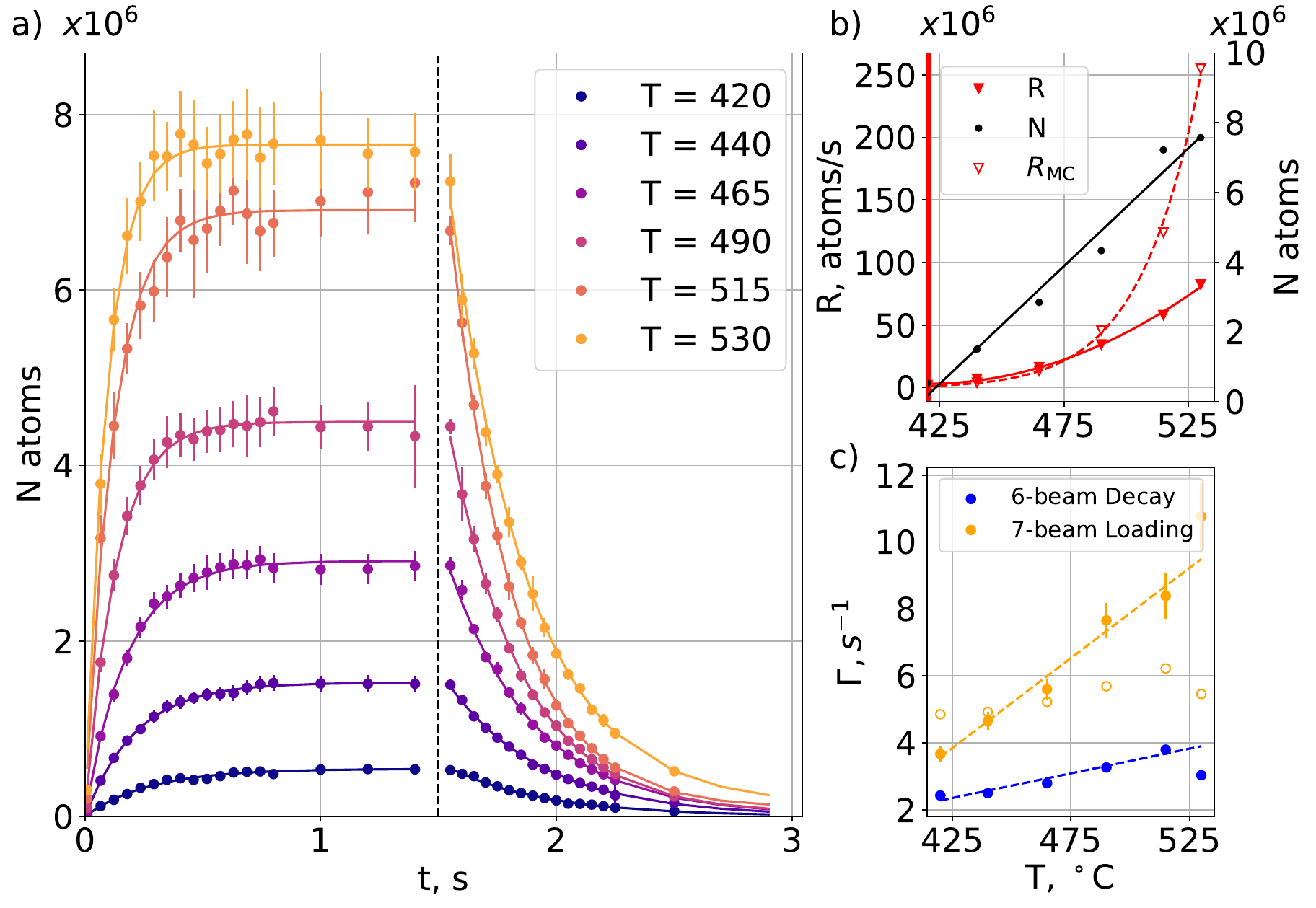}
\caption{\label{fig:oven} Blue MOT dynamics at the different oven temperatures. a) MOT atom number as a function of time. Vertical dashed line indicates that loading and decay processes were obtained via different measurements. b) Loading rate (left axis, red triangles), atoms number (right axis, black circles), and Monte-Carlo simulation loading rates (left axes, red triangles) as a function of oven temperature.}
\end{center}
\end{figure}

Another way to increase the loading rate and the number of atoms in the trap is to operate oven at a higher temperature. 
We measured the loading and unloading dynamics of the first-stage MOT for oven temperatures in the range from $420^\circ$\,C to $530^\circ$\,C (see Fig.\,\ref{fig:oven}.a).
Figure\,\ref{fig:oven}.b shows inferred loss rates during unloading (blue points, Zeeman beam is off and 6 MOT beams have $50\%$ of the initial power) and loading (orange filled points). 
Empty orange points depict expected loss rates during loading stage, which are obtained from unloading $\Gamma^l$ with addition of $\Gamma^\textrm{l,z}_{410}$ and scaling $\Gamma^\textrm{l,mot}_{410}$ for 2 times higher MOT beams power.
Origin of a noticeable discrepancy between them at higher oven temperatures is not clear yet. 
This, in addition to possible deviation of the velocity distribution in the hot atomic beam from capillary and inaccuracy in temperature readings, leads to a quadratic (filled red triangles, solid red lined) instead of expected exponential (empty red points, dashed red line) increase of the loading rate with the oven temperature, as it is shown in Fig.\,\ref{fig:oven}.c.
The equilibrium number of atoms (black points) increases linearly.

At this point we obtained loading rates up to $8\times10^8$ atoms/s and up to 8\,mln trapped atoms with a lifetime around 350\,ms.
We do not observe any kind of saturation, so these characteristics could be potentially increased.

For the next chapter, unless specified, we load 1\,mln atoms in the first-stage MOT and operate at the oven temperature $T = 465\,^\circ C$ at 80\,mW laser power.

\section{Cold atomic beam in the science chamber}\label{sec:atomic_beam}

Laser cooling of atoms in the primary chamber is the first step prior to loading of the magneto-optical trap and optical dipole trap in the science chamber.
The two-chamber configuration allows one to achieve high vacuum in the science chamber, which is not affected by the atomic oven and flux of hot atoms from it.
A new but already well established approach is to generate a cold atomic beam, directed to the science chamber MOT region, using a 2D-MOT in the primary chamber.

\begin{center}
\begin{figure}[h]
\includegraphics[width=\linewidth]{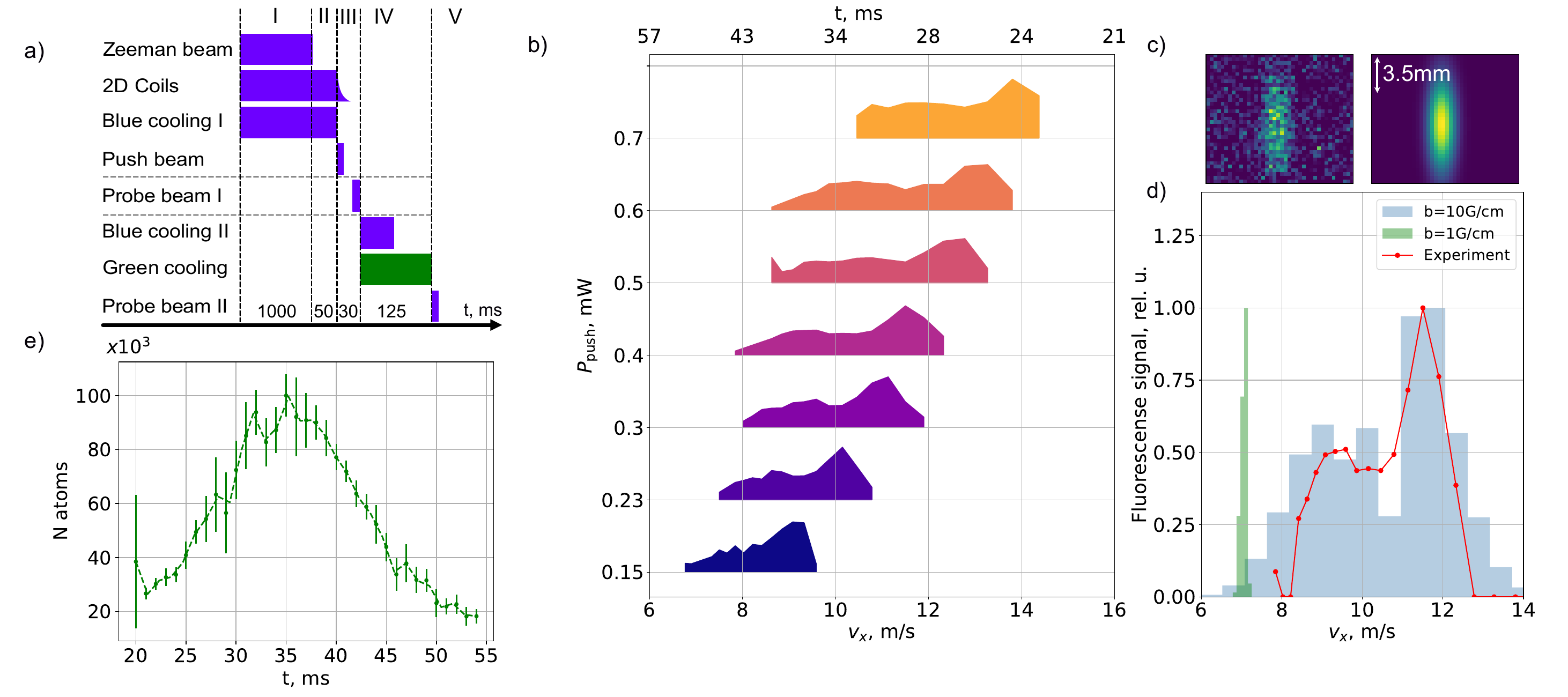}
\caption{\label{fig:Fig6long}
Spectroscopy of the cold atomic beam and second-stage MOT in the science chamber. a) Pulse sequence: I - first-stage MOT loading in the primary chamber, II - Atomic cloud compression, III - Push and free-flight stages, ``Probe beam I'' is used only for the atomic beam spectroscopy, IV - Atoms recapture in the science chamber MOT, V - Readout of the atoms number in the narrow-line MOT. Time axis is not to scale.
b) Velocity distributions in the cold atomic beam for different push beam powers. 
c) Example of a single image of the atomic beam (left) and 2D gaussian fit to it (right). 
d) Velocity distribution in cold atomic beam obtained with Monte-Carlo simulation in present configuration with magnetic field gradient $b=10$\,G/cm (gray bars) and for $b=1$\,G/cm (green bars); red points depict experimental data ($P_\textrm{push}=XX$\,mW). 
e) Number of atoms in the narrow-line MOT vs the length of stage III.}
\end{figure}
\end{center}

First, we investigated the properties of a cold beam of thulium atoms.
We start with accumulation of cold atoms in the primary chamber in the 7-beam configuration described above, then we turn off the Zeeman beam and wait for 50\,ms for the atoms to reach a steady-state position (see pulse sequence in Fig.\ref{fig:Fig6long}.a).
To reduce magnetic field gradient during the subsequent push-beam stage, we turn off the anti-Helmholtz coils 1\,ms before switching off the MOT beams. 
The magnetic field decay time constant is about $\tau_\textrm{coils} = 5$\,ms, so we can not eliminate the magnetic field gradient completely.
Then, we apply a 1.5\,ms push laser pulse to accelerate atoms towards the science chamber along the X-axis.
After a variable delay, we turn on the resonant probe beam ($P=5$\,mW, $\sigma=2$\,mm at $1/e^2$ intensity level), which is aligned vertically through the center of the science chamber (see sketch in Fig.\,\ref{fig:vacuum}.c), to detect the fluorescence of the cold atoms passing through.
Plots in Fig.\,\ref{fig:Fig6long}.b depict the measured luminescence signal (in relative units) as a function of the probe beam delay (top horizontal axis) for different powers of the push beam in the range from 0.15\,mW to 0.7\,mW.
Knowing the distance between the initial position of atoms in the primary chamber and the probe beam in the science chamber ($L=34.5$\,cm), we infer the velocity distribution in the cold atomic beam (bottom horizontal axis). 
The push beam diameter is $w_\textrm{push} = 2$\,mm, which ensures uniform power distribution over the initial cloud of atoms in the first-stage MOT in the primary chamber.
As expected, the average velocity of the atoms increases with push beam power from $\sim8$\,m/s to $\sim13$\,m/s.
In the meantime, we observed bimodal velocity distribution for all push beam powers.

In order to explain this effect we performed a Monte-Carlo simulation of the acceleration process \cite{yaushev2023two}. 
For this we calculated velocities acquired during a 1.5\,ms acceleration with a push-beam for 1000 atoms, which had initial Gaussian spatial distribution with $w=1$\,mm and Maxwell velocity distribution centered at $T = 200$\,\textmu K, corresponding to the Doppler limit for the 410\,nm transition.
In this model we implemented the real geometry of the anti-Helmholtz coils. 
Numerical simulations reproduced the experimental data (Fig.\,\ref{fig:Fig6long}.d, blue bars) and show that the bimodal structure is caused by the magnetic field gradient along the push beam direction.
In the case of near-to-zero magnetic field gradient simulations predict a single-peak narrow distribution of atomic velocities in the beam (Fig.\,\ref{fig:Fig6long}.d, green bars).
In the future we plan to switch to permanent magnets, which allow formation of magnetic field gradient along two axes and zero field along the third, enabling production of almost monochromatic atomic beam.

From the fluorescence images (Fig.\ref{fig:Fig6long}.c) we infer that the size of the atomic beam along the vertical direction is $w = 6.2$\,mm (FWHM), which corresponds to the angular spread of 18\,mrad of the initial beam after acceleration.

\section{Green MOT loading}\label{sec:green_mot}

For the second-stage MOT we have about 3\,mW of 530\,nm laser radiation available for this experiment.
Cooling beams are set to have the maximum allowed by the viewports $1/e^2$ radius of 8.5\,mm, which corresponds to the saturation parameter $s=2.2$.
While the cooling beams are larger than the measured atomic beam transversal dimension $w = 6.2$\,mm,  the capture efficiency can still be limited due to a moderate saturation intensity.

The upper bound for the capture velocity of the second-stage MOT is given by 
\begin{equation}\label{eq:vcap}
   \mathrm{v_{cap}} = \sqrt{\frac{\hbar k d \Gamma_{530} s}{2 m (1+s)}}\sim8\,\textrm{m/s},
\end{equation}
where $k = 2\pi / \lambda$ is the wave vector, $\lambda = 530$\,nm is the wavelength of the second stage cooling transition, $d=19$\,mm is the beam diameter, $\Gamma_{530}=350$\,kHz is the natural linewidth, $s=2.2$ is the saturation parameter, and $m=169$\,a.u is the Tm atomic mass.  
It is worth noting that for other lanthanides, e.g., Dy, Yb, the second-stage MOT capture velocity is of the same order, and additional techniques like core-shell MOT configuration \cite{PhysRevA.110.033103}, angled slowing beams \cite{lunden2020enhancing, Plotkin_Swing_2020} or side-band enhanced MOT \cite{PhysRevApplied.13.014013} are usually implemented to increase the capture efficiency into the narrow-line MOT.

Owing to a bimodal velocity distribution of atoms in the cold beam described above, high average velocity, and limited laser power, we did not observe loading of the second-stage MOT directly from the obtained cold atomic beam. 
However, we developed a different approach, where we use one of the blue cooling beams from the first-stage MOT, which is aligned with the cold atomic beam (see Fig.\,\ref{fig:vacuum}.c), to provide initial deceleration of atoms along the atomic beam direction.

We start by launching the atoms from the primary chamber with the same sequence as described in the previous chapter. 
After a variable delay $t=20-50$\,ms we simultaneously turn the second-stage MOT beams (530\,nm) and first-stage MOT beams (including the one along the X-axis) on to recapture atoms in the science chamber. 
After 75\,ms, we turn off the blue MOT beam and leave atoms trapped by only 530\,nm light for about 50\,ms to reach the steady-state atomic cloud position and temperature, and then image them using the vertical blue probe beam.
Measured number of atoms as a function of delay time $t$ is shown in Fig.\,\ref{fig:Fig6long}.e. 
There is an optimum delay time $t=35$\,ms, when we achieve trapping of $10^5$ atoms, corresponding to overall efficiency of $10\%$. 
For shorter $t$ blue cooling beam accelerates atoms in the beam, while for longer $t$ atoms fly over the trapping region.  
We found the best performance of the green MOT at the detuning $\Delta_g = -8\,\Gamma_{530}$ from resonance and magnetic field gradient around 3\,G/cm.

\begin{figure}[h!]
\includegraphics[width=0.5\linewidth]{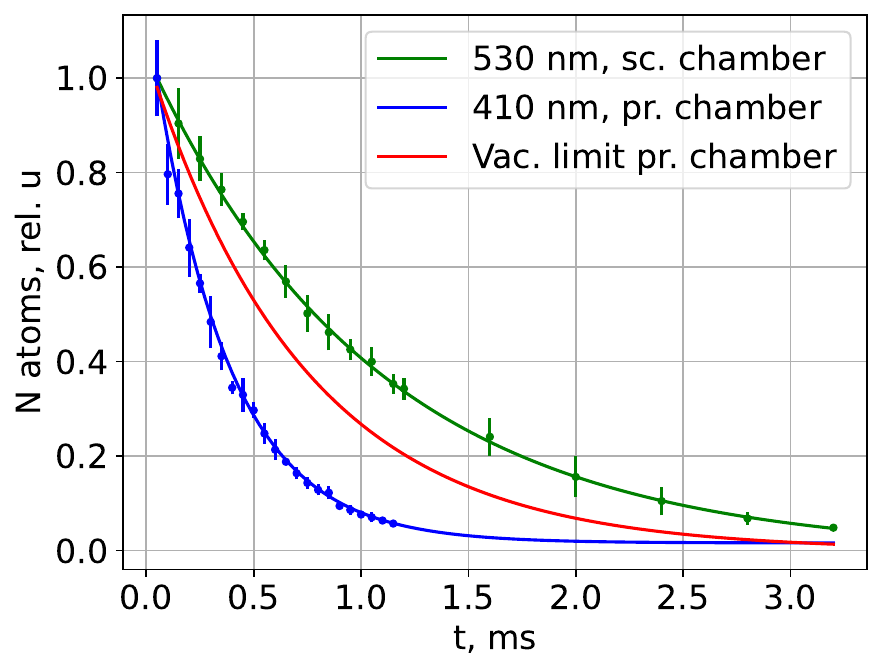}
\caption{\label{fig:lifetimes} Lifetimes comparison in the different chambers. Blue data indicates lifetime in the first-stage MOT in the primary chamber, green data corresponds to the lifetime in the second-stage MOT in the science chamber. Red line is the theoretical upper bound for the vacuum limited lifetime in the primary chamber given by $\Gamma^l_\textrm{coll}$ from Eq. \ref{eq:gamma_coll}.}
\end{figure}

Finally, with about $10^5$ atoms in the science chamber trapped in the narrow-line MOT, we were able to compare vacuum lifetimes in each chamber.
We measured the lifetime $\tau_g=1068(17)$\,ms of the atoms in the second-stage MOT in the science chamber (see Fig \ref{fig:lifetimes}, green points, solid green line is the exponential fit). 
This is significantly longer than the lifetime $\tau_b = 348(9)$\,ms in the first-stage MOT (blue points), and almost 50\% longer than the vacuum-limited lifetime $\tau^* = 731(11)$\,ms in the primary chamber (red line, for oven temperature $T=465\,^\circ$C).   
The first-stage MOT lifetime data is taken from Sec.\,\ref{sec:blue_mot}.C.

\section{Conclusion}\label{sec:end}

In this work, we have presented a two-chamber setup with a pulsed source of cold Tm atoms and loading of the narrow-line MOT in the science chamber from the cold atomic beam. 
In the primary chamber we have demonstrated MOT loading rate of more than $8\times10^7$\,atoms/s and more than $7\times10^6$ trapped atoms.
Operation in the pulsed mode allowed us to overcome difficulties from the broad velocity distribution in produced cold atomic beam due to a strong gradient of the MOT magnetic field in the primary chamber along the push beam axis.
We implemented a novel technique of using the horizontal beam from the primary first-stage MOT to enhance the capture velocity of the narrow-line MOT in the science chamber.
This proved to be crucial for the successful recapture of atoms into the second-stage MOT in our setup.
We achieved transfer efficiency of $\eta = 10\,\%$, which is primarily limited by the cold atomic beam velocity spread and available second stage laser power. 
We also observe the atoms lifetime of $\sim1$\,s in the science chamber, which is higher than in the primary chamber.
It worth noting that the vacuum level in the system was limited by poor gatewalve and reached $10^{-11}$\,mbar after its replacement. 

The loading rate in the primary chamber MOT could be increased by using more cooling radiation power as we observe linear grows up to the maximum available power in our setup.
Moreover, switching to the permanent magnet configuration for primary chamber MOT would drastically improve the cold atomic beam velocity homogeneity and deceleration efficiency by the Zeeman beam due to achievable larger magnetic field gradient along the direction of the hot atomic beam from the oven.
It would also allow realization of a 2D-MOT in the primary chamber, formation of continuous cold atomic beam and direct loading of the second-stage MOT in the science chamber.
In this mode one can achieve the highest number of atoms (more than $10^7$) as losses in the narrow-line MOT in the science chamber would be determined only by the vacuum lifetime.
On the other hand, operation in the pulsed mode opens the way to quickly refill MOT (within $\sim100$\,ms) in the science chamber from the primary chamber MOT, which is operated in parallel. 
This is particularly beneficial for experiments when a moderate number of atoms ($\sim10^6$) is sufficient, but high repetition rate (quantum computers and simulators) or short dead time (optical clocks) are essential. 
This method could be implemented in other systems with narrow-linewidth second-stage cooling transition (Sr, Yb, Er, and Dy).

\begin{acknowledgments}

The authors acknowledge the support of RSF grant no. 23-22-00437 . 

\end{acknowledgments}
 
\bibliography{aipsamp}%
\end{document}